\documentclass[aps,prb,superscriptaddress,twocolumn]{revtex4-2}
\usepackage{amsfonts, amssymb, amsmath, amsthm, graphicx}
\usepackage{mathrsfs}
\usepackage{braket}
\usepackage{subfigure}
\usepackage{multirow}
\usepackage{dcolumn}
\usepackage{bm}
\usepackage{color}
\usepackage[usenames,dvipsnames]{xcolor}
\usepackage[all]{xy}

\usepackage{array}
\usepackage{physics}

\renewcommand{\bf}[1]{\textnormal{\textbf{#1}}}
\newcommand{\BZ}{\mathrm{BZ}}

\newcommand{\meff}{m_{\rm eff}}

\usepackage[colorlinks=true,allcolors=blue]{hyperref}

\definecolor{Nathanblue}{rgb}{0.,0.24,0.51}

\newcommand{\blue}{\color{Nathanblue}}

\begin{document}

%\title{{\blue Perfect elliptic dichroism in anisotropic quantum Hall systems: \\ Probing  the metric of quantum Hall droplets}
%}

\title{{\blue Perfect elliptic dichroism: \\ Probing  the metric of anisotropic quantum Hall droplets}
}

\author{Bruno Mera}
%\textsuperscript{$\dagger$}}
\thanks{These authors contributed equally to this work.}
\affiliation{Instituto de Telecomunica\c{c}\~oes and Departmento de Matem\'{a}tica, Instituto Superior T\'ecnico, Universidade de Lisboa, Avenida Rovisco Pais 1, 1049-001 Lisboa, Portugal}
\affiliation{Advanced Institute for Materials Research (WPI-AIMR), Tohoku University, Sendai 980-8577, Japan}

\author{Alberto Nardin} %\textsuperscript{$\dagger$}}
\thanks{These authors contributed equally to this work.}
\affiliation{Universit\'{e} Paris-Saclay, CNRS, LPTMS, 91405, Orsay, France}

\author{Anaïs Defossez}
\affiliation{Laboratoire Kastler Brossel, Collège de France, CNRS, ENS-Université PSL,
Sorbonne Université, 11 Place Marcelin Berthelot, 75005 Paris, France}
\affiliation{International Solvay Institutes, B-1050 Brussels, Belgium}
\affiliation{Center for Nonlinear Phenomena and Complex Systems,
Université Libre de Bruxelles, CP 231, Campus Plaine, B-1050 Brussels, Belgium}

\author{Baptiste~Bermond}
\affiliation{Laboratoire Kastler Brossel, Collège de France, CNRS, ENS-Université PSL,
Sorbonne Université, 11 Place Marcelin Berthelot, 75005 Paris, France}

\author{Tomoki Ozawa}
\affiliation{Advanced Institute for Materials Research (WPI-AIMR), Tohoku University, Sendai 980-8577, Japan}
\affiliation{RIKEN Center for Interdisciplinary Theoretical and Mathematical Sciences (iTHEMS), RIKEN, Wako, Saitama 351-0198, Japan}

\author{Nathan Goldman}
\email{nathan.goldman@lkb.ens.fr}
\affiliation{Laboratoire Kastler Brossel, Collège de France, CNRS, ENS-Université PSL,
Sorbonne Université, 11 Place Marcelin Berthelot, 75005 Paris, France}
\affiliation{International Solvay Institutes, B-1050 Brussels, Belgium}
\affiliation{Center for Nonlinear Phenomena and Complex Systems,
Université Libre de Bruxelles, CP 231, Campus Plaine, B-1050 Brussels, Belgium}
%\date{}
%
\begin{abstract}
Understanding the geometry of quantum Hall systems is a central challenge in modern condensed matter physics. We introduce a framework for probing the geometric structure of quantum Hall droplets by engineering the geometry of a dichroic probe and identifying the onset of ``perfect elliptic dichroism", a regime in which the system responds exclusively to an elliptically polarized drive of a given chirality. This phenomenon provides a direct diagnostic of the droplet's intrinsic metric, and we show that it extends naturally to ideal Chern bands, where holomorphicity of the occupied states guarantees the vanishing of one chiral absorption rate with a quantized response for the other. In lattice realizations, such as the Harper–Hofstadter model, finite lattice-spacing corrections break the exact continuum metric description and give rise to a renormalized, emergent Landau-orbit metric; the probe ellipticity at which perfect dichroism is achieved then shifts accordingly, offering a direct spectroscopic window onto this lattice-induced geometric renormalization. Our results illuminate the rich geometric structure of quantum Hall phases and offer concrete pathways for observing these effects in quantum-engineered platforms.
\end{abstract}
\maketitle 

\section{Introduction}
\label{sec:introduction motivations and key contributions}
Quantum Hall systems serve as a fertile ground for exploring the geometric foundations of quantum matter. In both continuum and lattice realizations, these systems exhibit rich internal structures governed by concepts from differential geometry and complex analysis. The geometric viewpoint on quantum Hall states was emphasized in the context of fractional quantum Hall physics~\cite{Haldane_2011}, while Kähler geometry, complex structures, and quantum metric tensor formulations have become central to understanding Chern bands and their quantum geometry~\cite{mera:ozawa:2021,mera:ozawa:2021:engineering,liu:mera:fujimoto:ozawa:wang:2025,paiva:wang:ozawa:mera:2025, xavier2025prl,yu2025quantum}. These geometric quantities manifest directly in physical observables, such as the Hall conductivity and orbital magnetic response, where the Berry curvature, the quantum metric, and related quantities contribute to measurable transport coefficients~\cite{nagaosa2010rmp,xiao2010rmp,yu2025quantum,gao2025quantum,souza2008prb,resta2020prr}. Synthetic lattice systems, engineered in moir\'e materials, photonics or cold-atom setups, further enrich this geometrical landscape by enabling tunable topology and enhanced interaction effects~\cite{ozawa:price:amo:goldman:etal:2019,cooper:dalibard:spielman:2019,Nascimbene_Dalibard,sommer2025idealopticalfluxlattices,wang2022prl,ledwith2022prl}.

To access these geometric properties experimentally, sum rules and circular dichroism have emerged as incisive tools, linking optical responses and excitation rates under periodic driving to the underlying quantum geometry. Sum rules impose rigorous constraints on integrated response functions, enabling the systematic extraction of Berry-curvature and quantum-metric contributions from optical absorption spectra~\cite{souza2000prb,souza2008prb,tran:dauphi:grushin:zoller:goldman:2017,ozawa:goldman:2018,ozawa2019prr,resta2020prr,kruchkov2023spectral,goldman2024relating,onishi2024fundamental,onishi2024topological,bermond2025dichroism,kruchkov2025bounds,ji2025density,shinada2025quantum}. Circular dichroism, which measures the differential absorption of left- and right-circularly polarized drives, provides a particularly direct handle on time-reversal and inversion-symmetry breaking, giving experimental access to topological invariants such as the Chern number, both in materials~\cite{beaulieu2024berry,kang2025measurements,yu2024quantum} and in synthetic quantum systems~\cite{asteria2019measuring,tan2019experimental,yu2020experimental,klees2020microwave,chen2022synthetic}. Together, these approaches form a unifying experimental paradigm for investigating quantum geometry across disparate platforms, from two-dimensional electron gases and moiré materials to ultracold atoms in optical lattices and photonic systems.

Among two-dimensional systems exhibiting quantized Hall conductance, ideal Chern bands form a distinguished class with particularly rich physical and mathematical properties~\cite{ozawa:mera:2021,mera:ozawa:2021,mera:ozawa:2021:engineering,wang:cano:millis:liu:yang:2021,wang:klevstov:liu:2023,wang2022prl,ledwith2022prl,ledwith:vishwanath:parker:2023,mera:ozawa:2024,liu:mera:fujimoto:ozawa:wang:2025,Nascimbene_Dalibard,sommer2025idealopticalfluxlattices,cano:wang:2026}. Physically, such systems are expected to stabilize fractional quantum Hall or fractional Chern insulating states upon the introduction of suitable inter-particle interactions, making them prime candidates for realizing strongly correlated topological phases. The lowest Landau levels provide the paradigmatic continuum examples of ideal Chern bands. Mathematically, the Bloch states of an ideal Chern band can be chosen to be holomorphic functions of the crystal momentum; as a consequence, the two-dimensional Brillouin zone, equipped with the quantum metric and Berry curvature, acquires the structure of a Kähler manifold~\cite{mera:ozawa:2021,ozawa:mera:2021,mera:ozawa:2021:engineering,wang:cano:millis:liu:yang:2021,wang:klevstov:liu:2023,mera:ozawa:2024,liu:mera:fujimoto:ozawa:wang:2025,cano:wang:2026}. It is precisely this holomorphic structure that, as we will show, underlies the phenomenon of perfect elliptic dichroism.

In this paper, we study quantum Hall droplets of arbitrary shape through their elliptic dichroic response. The anisotropic shape of a quantum Hall droplet can originate from an anisotropic effective mass, which enters the kinetic energy through an anisotropic metric and deforms the cyclotron orbits from circles into ellipses~\cite{Haldane_2011,avron:seiler:zograf:1995,tokatly2007prb,bo2012prb}. We show that the lowest Landau level, and more generally any filled ideal Chern band, exhibits a remarkably sharp response to an elliptically polarized drive whose geometry is adapted to this intrinsic metric. Specifically, when the ellipticity of the probe matches the anisotropy of the effective mass, the absorption rate for one chirality of the drive vanishes exactly, while the rate for the opposite chirality is quantized and proportional to the integer Chern number. This all-or-nothing chiral response stands in sharp contrast to the situation for generic Chern bands, where neither rate vanishes individually and only their \emph{difference} is proportional to the Chern number~\cite{tran:dauphi:grushin:zoller:goldman:2017,asteria2019measuring,Repellin_dichroism,goldman2024relating}. It also generalizes the notion of perfect circular dichroism, previously established for isotropic ideal-band systems~\cite{junkai:wang:ledwith:vishwanath:parker:2023}, to the anisotropic setting. We call this phenomenon \textit{perfect elliptic dichroism}, and propose it as a direct experimental diagnostic of the intrinsic metric geometry of quantum Hall droplets. A related notion has been discussed in the context of anisotropic Dirac and magnetic systems~\cite{ezawa:2014,ezawa:2025,ezawa:2026:a,ezawa:2026:b}, where matching the probe ellipticity to the local anisotropy of a given, gapless Dirac cone enables valley-selective optical pumping. The mechanism underlying our result is of a fundamentally different nature: rather than a local effect tied to the dispersion of individual Dirac cones, perfect elliptic dichroism here is a global, topologically quantized response of a gapped Chern band, rooted in the holomorphic (K\"ahler) structure of the Bloch states of ideal Chern bands and of fractional quantum Hall states.

A quantity intimately related to the geometrical structure of quantum Hall droplets is the Hall viscosity, the nondissipative stress response to adiabatic deformations of the underlying spatial metric~\cite{avron:seiler:zograf:1995,tokatly2007prb}, which is closely connected to the intrinsic metric geometry of quantum Hall droplets~\cite{Haldane_2011}. The formalism developed in this paper lays out the foundations for the accompanying paper~\cite{sisterViscositypaper}, in which we develop metric-sensitive dichroic probes to measure the Hall viscosity.

The paper is organized as follows. In Sec.~\ref{sec:the metric of quantum hall settings} we review the geometric description of quantum Hall systems, with particular emphasis on the role of the anisotropic effective mass and the associated complex structure underlying the Landau level algebra; we also discuss how anisotropic Landau levels can be obtained from isotropic ones via a squeezing transformation. In Sec.~\ref{sec: elliptic dichroic response} we introduce the elliptic dichroic probe and establish the phenomenon of perfect elliptic dichroism: for ideal Chern bands, the holomorphicity of the occupied states forces one chiral absorption rate to vanish exactly, while the other is fixed by the Chern number. We further show that this result extends to fractional quantum Hall states via a many-body flux-space formulation. In Sec.~\ref{sec:application} we benchmark these predictions against exact diagonalization of the Harper–Hofstadter model, demonstrating that lattice corrections shift the optimal probe ellipticity away from its continuum value and interpreting this shift as the signature of a renormalized, emergent Landau-orbit metric. Finally, Sec.~\ref{sec:outlook} summarizes our findings and discusses future directions.

\section{The metric of Quantum Hall settings}
\label{sec:the metric of quantum hall settings}
When considering the single-particle quantum mechanical problem of electrons on a two-dimensional (oriented) surface under a perpendicular magnetic field, one must geometrically specify two ingredients: a metric tensor $g_{ab}$ and a gauge field $A_a$. Once the metric is fixed, the notion of a uniform magnetic field---understood as the curvature $F_{ab}=\partial_a A_b-\partial_b A_a$ of the gauge field---makes sense in a coordinate-independent way: $F_{ab}=B\sqrt{\det(g)}\epsilon_{ab}$, for some constant $B$. Here $\sqrt{\det(g)}\epsilon_{ab}$ is the area $2-$form associated with the metric tensor $g_{ab}$, and $\epsilon_{ab}=\epsilon^{ab}$ is the Levi-Civita symbol in two dimensions, defined by $\epsilon_{12} = -\epsilon_{21} = 1$ and $\epsilon_{11} = \epsilon_{22} = 0$~\footnote{Note that the local expression $\sqrt{\det(g)}\epsilon_{ab}$ for the area $2$-form assumes the local charts are positively oriented.}. Usually, one typically assumes that the surface is the plane equipped with the Euclidean metric and avoids such geometric considerations. However, to discuss response functions such as the Hall conductivity and the Hall viscosity, the gauge field and the metric play a crucial role. Indeed, both responses admit a unified description within adiabatic response theory:~the Hall conductivity arises from threading a flat gauge field through the system -- parametrized by twist angles and leaving the magnetic field unchanged~\cite{goldman2024relating} -- while the Hall viscosity arises from performing area-preserving variations of the flat metric, as established in the seminal work of Avron, Seiler, and Zograf~\cite{avron:seiler:zograf:1995}. In both cases, the response coefficient is identified as a Berry curvature in the space of external parameters.

Below we provide a self-contained review of the geometric description of Landau levels in an anisotropic setting. Readers interested in a more detailed geometric discussion of issues related to Hall viscosity may consult Ref.~\cite{read:rezayi:2011}.

We restrict our study to flat, spatially uniform metrics in order to isolate homogeneous, area-preserving anisotropic deformations of the Landau problem, without introducing spatial curvature effects. Concretely, the Landau Hamiltonian on the plane equipped with a flat unimodular metric $g = g_{ab}dr^a dr^b$, with $\sqrt{\det(g)}=1$, and uniform magnetic field $B>0$ is given by 
\begin{align}
H=\frac{1}{2m} g^{ab}\pi_a\pi_b, 
\label{eq: anisotropic LL Hamiltonian}
\end{align}  
where $m$ is the mass, $g^{ab}$ is the inverse of $g_{ab}$ and $\pi_a$ are the kinetic momenta defined by  
\begin{align}
\pi_a = p_a -eA_a, 
\end{align}  
where $A_a$ is the magnetic vector potential and $e$ the electric charge. Throughout this work, we adopt the Einstein summation convention. In this paper, we use notation for the coordinates $(r^1, r^2)$ and $(x,y)$ interchangeably; they are equivalent. The effect of the anisotropic metric $g^{ab}$ is to deform the cyclotron orbits from circles into ellipses. Physically, the anisotropy can arise either from uniform shear stress applied to an originally isotropic sample, from the presence of a tilted magnetic field in samples with finite layer thickness, or from the anisotropic effective mass due to the underlying lattice anisotropy; see Ref.~\cite{ando:1982,avron:seiler:zograf:1995,Goldman_PRL_2009,yang:rezayi:bhatt:haldane:2012,Oblak_PRX}.

\subsection{Geometric description of Landau levels}
\label{subsec:geometric description of landau levels}
The kinetic momenta satisfy the commutation relations  
\begin{align}
\label{eq:pis_commutator}
[\pi_a, \pi_b]=i \hbar e F_{ab}= i \hbar e B \epsilon_{ab}.
\end{align}  
The above commutation relations follow from the statement that the magnetic field is the curl of the magnetic vector potential $\partial _aA_b -\partial _bA_a=B\epsilon_{ab}$. 

We introduce creation and annihilation operators, 
\begin{align}
a_{\tau} &=\frac{1}{\sqrt{\hbar e B}}\omega^a \pi_{a}=\frac{1}{\sqrt{2\tau_2 \hbar e B}}(\pi_1+\tau \pi_2), \nonumber \\
a^\dagger_{\tau}&=\frac{1}{\sqrt{\hbar e B}}\omega^{*a}\pi_a=\frac{1}{\sqrt{2\tau_2 \hbar e B}}(\pi_1+\tau^*\pi_2),
\label{eq: complex structure tau orbital operators}
\end{align}
with complex parameters $\tau = \tau_1 + i \tau_2$ and $(\omega^1, \omega^2) \equiv (\sqrt{2\tau_2})^{-1}(1,\tau)$. We note that
\begin{align}
[a_{\tau},a_{\tau}^\dagger]=1.  \end{align}
The complex parameter $\tau$, with a condition $\tau_2 > 0$, will be chosen so that $H$ is brought into a harmonic oscillator form:
\begin{align}
H=\frac{1}{2}\hbar\Omega \left(a_{\tau}^\dagger a_{\tau} + a_{\tau}a_{\tau}^\dagger\right).
\end{align}
Explicitly, 
\begin{align}
&\frac{1}{2}\hbar \Omega  \left(a_{\tau}^\dagger a_{\tau} + a_{\tau}a_{\tau}^\dagger\right) =\frac{\Omega}{e B}\; \mathrm{Re}(\omega^{*a} \omega^b) \pi_a\pi_b \nonumber \\
&=\frac{1}{2}\frac{\Omega}{\tau_2 e B}\left[\pi_1^2 +\tau_1\left(\pi_1\pi_2 +\pi_2\pi_1\right)+ |\tau|^2 \pi_2^2\right].
\end{align}
Comparing with Eq.~\eqref{eq: anisotropic LL Hamiltonian}, we see that the complex parameter $\tau$ is related to $g^{ab}$ by
\begin{align}
\begin{bmatrix}
g^{ab}    
\end{bmatrix}=\begin{bmatrix}\frac{m \Omega}{\tau_2 q B}\; 2\mathrm{Re}(\omega^{*a}\omega^b)
\end{bmatrix}=\frac{m\Omega}{\tau_2 e B}\begin{bmatrix}
1 & \tau_1 \\
\tau_1 & |\tau|^2
\end{bmatrix},
\label{eq: unimodular anisotropic flat metric}
\end{align}
and from unimodularity of $g$, we get the cyclotron frequency $\Omega=\frac{e B}{m}$. Changing $g$  while constraining it to be unimodular, we are probing the response of the system to shear stress, whose associated response coefficient is the Hall viscosity~\cite{avron:seiler:zograf:1995}. The parameter $\tau$ describes a choice of complex structure in the plane determined by the complex coordinate $z_{\tau}=-y +\tau x$, and the Hall viscosity is naturally interpreted as the Berry curvature associated to the change of the associated quantum states with respect to the change of this complex structure.

We now introduce the guiding center operators
\begin{align}
\label{eq:guiding_centers}
R^a= r^a -(e B)^{-1} \epsilon^{ab}\pi_b,
\end{align}
which satisfy the commutation relations
\begin{align}
[R^a,R^b]=-i\frac{\hbar}{ e B}\epsilon^{ab} =-i \ell^2_{B} \epsilon^{ab}.
\end{align}
They commute with the kinetic momenta $[\pi_a,R^b]=0$, and generate the magnetic translations commuting with $H$ in Eq.~\eqref{eq: anisotropic LL Hamiltonian}. 
Thus, the guiding center operators provide an independent harmonic oscillator algebra in addition to the kinetic momenta. We introduced the square magnetic length $\ell^2_B=\hbar/e B$. We can then introduce associated creation and annihilation operators
\begin{align}
&b_{\tau}= \sqrt{\frac{e B}{\hbar}}\; \omega^*_aR^a =\sqrt{\frac{e B}{2\tau_2\hbar}}\left(-R^2 +\tau^* R^1\right), \nonumber \\
&b_{\tau}^\dagger=\sqrt{\frac{e B}{\hbar}}\; \omega_a R^a= \sqrt{\frac{e B}{2\tau_2\hbar}}\left(-R^2 +\tau R^1\right) 
\label{eq: complex structure tau guiding center operators}
\end{align}
with $\omega_{a}=-i \epsilon_{ab}\omega^b$ the dual basis of $\omega^a$, so that $[b_{\tau},b_{\tau}^\dagger]=1$.  

We can then describe the Landau level states in terms of a Fock basis 
\begin{align}
\label{eq:fock_states}
\ket{\Psi_{n,l}(\tau)}=\frac{1}{\sqrt{n!}}\frac{1}{\sqrt{l!}}(a_{\tau}^\dagger)^n (b_{\tau}^\dagger)^l\ket{0},
\end{align}
where we explicitly denoted the $\tau$ dependence of the states above.

As in the isotropic case, which we recover setting $\tau=i$, $n$ labels the Landau level index and $l$ the number of powers of $z_{\tau}=-y+\tau x$ in a holomorphic gauge. Indeed, acting on $n=0$ states, the operator $b_{\tau}^\dagger$ is in that gauge, up to a constant, multiplication by $z_{\tau}$.
\subsection{Anisotropic Landau levels in terms of squeezing}
\label{subsec:anisotropic landau levels in terms of squeezing}
From the case $\tau=i$, which corresponds to the standard isotropic metric $\delta^{ab}$, we can go to the arbitrary anisotropic case described by $g^{ab}$ in Eq.~\eqref{eq: unimodular anisotropic flat metric}, by performing a Bogoliubov-Valatin squeezing transformation taking the operators $a_{\tau=i}$ and $a_{\tau=i}^\dagger$ to new operators $a_{\tau}$ and $a_{\tau}^\dagger$, for any $\tau$ in the upper half plane.

To see this, consider a single bosonic mode described by creation and annihilation operators $a^\dagger$ and $a$, satisfying the standard bosonic canonical commutation relations. Consider then a linear transformation of the bosonic operators of the form
\begin{align}
\begin{bmatrix}
\widetilde{a} & \widetilde{a}^{\dagger}    
\end{bmatrix}=\begin{bmatrix}
a & a^\dagger    
\end{bmatrix} \begin{bmatrix}
\alpha & \beta^*\\
\beta & \alpha^*
\end{bmatrix}.
\end{align}
This transformation preserves the commutation relations if and only if $|\alpha|^2-|\beta|^2=1$. That is to say that the matrix $U=\begin{bmatrix}
\alpha & \beta^*\\
\beta & \alpha^*
\end{bmatrix}\in SU(1,1)$. 
Observe that the vacuum and the states above it, which we can build by acting with the creation operator, are not affected by the change
\begin{align}
\begin{bmatrix}
\widetilde{a} & \widetilde{a}^{\dagger}    
\end{bmatrix}\longrightarrow \begin{bmatrix}
\widetilde{a} & \widetilde{a}^{\dagger}    
\end{bmatrix}\begin{bmatrix}
e^{i\phi} & 0\\
0 & e^{-i\phi}
\end{bmatrix},    
\end{align}
with $e^{i\phi}\in U(1)$. It follows that, from the physical point of view, we only care about $U\in SU(1,1)$ modulo $U(1)$, i.e., we are interested in the quotient space $M=SU(1,1)/U(1)$. Elements of $M$ can be parameterized by a complex vector $\begin{bmatrix}
\alpha & \beta    
\end{bmatrix}$ satisfying $|\alpha|^2-|\beta|^2=1$, modulo multiplication by a phase. Because of the $U(1)$-gauge redundancy, we can parametrize $M$ by the complex variable $w=\beta/\alpha$, provided $\alpha\neq 0$. However, since $|\alpha|^2-|\beta|^2=1$, we have that $|\alpha|^2\geq 1$, which then implies that $\alpha$ is never zero, so the parametrization is global. Furthermore, the same equation implies $1- |w|^2=1/|\alpha|^2$, which then yields $|w|^2=1-\frac{1}{|\alpha|^2}$, so that $|w|<1$, i.e. $w$ is in the unit disk $D=\{w\in \mathbb{C}:|w|<1\}$. We can parametrize $\alpha =\cosh(r)$ and $\beta=\sinh(r) e^{i\phi}$ and 
\begin{align}
w=\tanh(r)e^{i\phi}.
\end{align}
If we introduce a complex variable $\zeta$, we can arrange for $r=|\zeta|$ and $e^{i\phi}=\zeta/|\zeta|$, so that 
\begin{align}
U=\begin{bmatrix}
\cosh(|\zeta|) & \sinh(|\zeta|)\frac{\zeta^*}{|\zeta|}\\
\sinh(|\zeta|)\frac{\zeta}{|\zeta|} & \cosh(|\zeta|),
\end{bmatrix}
\end{align}
and we understand that $g$ comes from the standard squeezing operator $S(\zeta)=\exp\left[\frac{1}{2}\left(\zeta a^{\dagger 2} -\zeta^*a^{2}\right)\right]$, i.e.
\begin{align}
S(\zeta)\begin{bmatrix}
a & a^{\dagger}    
\end{bmatrix}S(\zeta)^{-1}=\begin{bmatrix}
\widetilde{a} & \widetilde{a}^{\dagger}    
\end{bmatrix}=\begin{bmatrix}
a & a^\dagger    
\end{bmatrix} \begin{bmatrix}
\alpha & \beta^*\\
\beta & \alpha^*
\end{bmatrix}.
\end{align} 
Now the disk $D$ is, as a complex manifold, isomorphic to the upper half space $\mathbb{H}=\{\tau=\tau_1+i\tau_2\in\mathbb{C}:\tau_2>0\}$, with the isomorphism given by the Cayley transformation
\begin{align}
w= \frac{1+i\tau}{1-i\tau},  
\end{align}
and inverse
\begin{align}
\tau=i\frac{1-w}{1+w}.
\end{align}
Because
\begin{align}
\tau_2=\frac{1-|w|^2}{|1+w|^2}>0 ,\text{ given } |w|<1,
\end{align}
it is clear that $i(1-w)/(1+w)\in \mathbb{H}$ and hence, because we have a holomorphic map with holomorphic inverse, this is an isomorphism of complex manifolds. We can then read off the parameters of the squeezing operator to achieve arbitrary $\tau$:
\begin{align}
w=\tanh(|\zeta|)\frac{\zeta}{|\zeta|}=   \frac{1+i\tau}{1-i\tau}. 
\end{align}
Recalling Eq.~\eqref{eq: complex structure tau orbital operators}, we can write
\begin{align}
\label{eq: atau in terms of ai}
&\begin{bmatrix}
a_{\tau} & a_{\tau}^{\dagger}    
\end{bmatrix}=\frac{1}{\sqrt{2\tau_2 \hbar eB}}\begin{bmatrix}
\pi_1 & \pi_2    
\end{bmatrix}\begin{bmatrix}
1 & 1\\
\tau & \tau^*
\end{bmatrix} \nonumber \\
&=   \frac{1}{2\sqrt{\tau_2}}\begin{bmatrix}
a_i & a_i^\dagger    
\end{bmatrix}\begin{bmatrix}
1 & -i\\
1 & i
\end{bmatrix}\begin{bmatrix}
1 & 1\\
\tau & \tau^*
\end{bmatrix} \nonumber \\
&=\frac{1}{2\sqrt{\tau_2}}\begin{bmatrix}
a_i & a_i^\dagger    
\end{bmatrix}\begin{bmatrix}
1-i\tau & 1-i\tau^*\\
1+i\tau & 1+ i\tau^*
\end{bmatrix},
\end{align}
and we conclude that the squeezing parameter is $w=(1+i\tau)/(1-i\tau)$, as before. It then follows that
\begin{align}
&S(\zeta)\left(\frac{1}{2m}\delta^{ab}\pi_a \pi_b\right)S^{-1}(\zeta)\nonumber \\
&=S(\zeta)\left[\frac{1}{2}\hbar\Omega\left(a_{i}^\dagger a_i +a_ia_i^{\dagger}\right)\right]S^{-1}(\zeta)\nonumber\\
&=\frac{1}{2}\hbar\Omega\left(a_{\tau}^\dagger a_\tau +a_\tau a_\tau^{\dagger}\right) \nonumber \\
&=  \frac{1}{2m}g^{ab}\pi_a\pi_b,
\end{align}
so the isotropic metric goes to an anisotropic metric, $\delta^{ab}\to g^{ab}$, under squeezing, as announced. We point out that Ref.~\cite{Goldman_PRL_2009} analyzed a similar relation between squeezing and anisotropic quantum Hall states in the context of relativistic Landau levels.
\section{Elliptic dichroic response}
\label{sec: elliptic dichroic response}
We now turn to the central question of this paper: how to probe the geometric and topological properties of a quantum Hall fluid using an elliptically polarized drive. Our primary focus is the fully filled lowest Landau level ($n=0$), which, as discussed in Sec.~\ref{sec:the metric of quantum hall settings}, is naturally an ideal Chern band with respect to the complex structure determined by $\tau$. The key insight is that the complex structure also defines a preferred elliptic polarization for the probe: when the probe geometry is chosen to be consistent with $\tau$, the holomorphicity of the lowest Landau level states forces the excitation rate for one chirality of the drive to vanish exactly. Physically, this vanishing reflects the fact that the probe has been perfectly adapted to the intrinsic anisotropy of the quantum Hall droplet: it ``sees'' the droplet's geometry and excites only the chirality that is compatible with it.

Driving the system elliptically can be conveniently described by adding a periodic electric field perturbation of the form
\begin{align}
\label{eq: dichroic probe}
V_{\pm}(t):=-e E^{\pm}_{a}(t)r^a=- \frac{2\mathcal{E}}{\sqrt{\tau'_2}}  \; \mathrm{Re}\left(z_{\tau'} e^{\mp i(\omega t+\delta)}\right), 
\end{align}
where $z_{\tau'}=-y+\tau' x$ , for $\tau'=\tau'_1+i\tau'_2$ not necessarily equal to $\tau$, and $\pm$ refers to the two different possible orientations of the driving. Additionally, $2\mathcal{E}$ is the overall real-valued amplitude, $\omega$ is the driving frequency and $\delta$ is a phase shift, which is introduced for later convenience. The elliptic driving is described by a periodic electric field $E^{\pm}_a(t)$, explicitly given by $(E^{\pm}_a(t))=(2\mathcal{E}/e\sqrt{\tau'_2})(\tau'_1\cos(\omega t)\pm \tau'_2\sin(\omega t),-\cos(\omega t))$, with constant magnitude $g'^{ab}E^{\pm}_{a}(t)E^{\pm}_{b}(t)=4\mathcal{E}^2/e^2$ with respect to the metric $g'^{ab}$ determined by $
\tau'$ as in Eq.~\eqref{eq: unimodular anisotropic flat metric}. The choice $\tau=\tau'$ plays a distinguished role:~it corresponds to adapting the ellipticity of the probe to the intrinsic metric of the quantum Hall droplet, and we will see that it is precisely this matching condition that causes one of the chiral excitation rates to vanish exactly for the $n=0$ level.

Using the relations
\begin{align}
x=R^1 +(e B)^{-1}\pi_2  \text{ and } y= R^2 -(e B)^{-1}\pi_1,  
\end{align}
one readily finds that
\begin{align}
\label{eq:probe_decomposition}
z_{\tau'}=\sqrt{\frac{2\tau'_2 \hbar}{e B}} \left(b^\dagger_{\tau'}+a_{\tau'}\right)  \text{ and } z^*_{\tau'}=\sqrt{\frac{2\tau'_2 \hbar}{e B}} \left(b_{\tau'}+a^{\dagger}_{\tau'}\right),
\end{align}
where the creation and annihilation operators are defined in the same way as before, just with $\tau$ replaced by $\tau'$. By the discussion in Sec.~\ref{subsec:anisotropic landau levels in terms of squeezing}, there exist matrices $U_1,U_2\in SU(1,1)$, corresponding to squeezing transformations, such that
\begin{align}
\begin{bmatrix}
a_{\tau'}  & a_{\tau'}^{\dagger}    
\end{bmatrix}    =\begin{bmatrix}
a_{\tau}  & a_{\tau}^{\dagger}    
\end{bmatrix} U_1 \text{ and } \begin{bmatrix} b_{\tau'}  & b_{\tau'}^{\dagger}    
\end{bmatrix}    =\begin{bmatrix}
b_{\tau}  & b_{\tau}^{\dagger}    
\end{bmatrix} U_2.
\end{align}
It follows that $z_{\tau'}$ will, in general, be a linear combination of $a_{\tau}, a_{\tau}^{\dagger},b_{\tau}$ and $b_{\tau}^{\dagger}$ (see Eq.~\eqref{eq: ztauprime in terms of btau,btaudagger,atau,ataudagger} for explicit form of the coefficients).

According to Fermi's golden rule, the frequency-integrated excitation rates ($\Gamma^{\mathrm{int}}_{\pm}=\int_{0}^{\infty}d\omega \; \Gamma_{\pm}(\omega)$), \emph{cf.}~\cite{tran:dauphi:grushin:zoller:goldman:2017, ozawa:goldman:2018}, can be written in the form 
\begin{align}
\label{eq: dichroic rates}
&\Gamma^{\mathrm{int}}_{+} \!=\! \frac{2\pi  \mathcal{E}^2}{\tau'_2\hbar^2}\sum_{n_2=1}^{\infty}\sum_{l_2=0}^{\infty}\sum_{l_1=0}^{l_F-1} \left|\bra{\Psi_{n_2,l_2}(\tau)} z_{\tau'}\ket{\Psi_{n_1=0,l_1}(\tau)}\right|^2, \nonumber \\
&\Gamma^{\mathrm{int}}_{-} \!=\! \frac{2\pi  \mathcal{E}^2}{\tau'_2\hbar^2}\sum_{n_2=1}^{\infty}\sum_{l_2=0}^{\infty}\sum_{l_1=0}^{l_F-1} \left|\bra{\Psi_{n_2,l_2}(\tau)} z^*_{\tau'}\ket{\Psi_{n_1=0,l_1}(\tau)}\right|^2.
\end{align}
Note that we have removed from the sums defining $\Gamma^{\mathrm{int}}_{\pm}$ those contributions that correspond to transitions within the lowest Landau level; these neglected contributions occur near the edge, namely from $(n_1=0,l_1=l_{F}-1)$ to $(n_2=0,l_2=l_{F})$, and cancel exactly the bulk contribution -- a manifestation of the bulk-to-boundary principle~\cite{tran:dauphi:grushin:zoller:goldman:2017,TranCooperGoldman18_PRA_circdichroLL}. In practice, we can neglect these contributions by taking $\omega$ greater than the bulk gap~\cite{unal_2025}. Under these conditions the area is defined by $A=2\pi \ell_B^2 l_F$ with $\ell_B^2= \hbar/(e B)$. In this lowest Landau level setup, the number of particles $N$ is simply linked with the Fermi point $l_F-1$ as $N=l_F$.
\subsection{Quantized circular dichroic response of an elliptic quantum Hall droplet}
\label{subsec:quantized circular dichroic response}
It follows from the results above [Eq.~\eqref{eq: dichroic rates}] that the relevant matrix elements are of the form
\begin{align}
&\bra{\Psi_{n_2,l_2}(\tau)} z_{\tau'}\ket{\Psi_{n_1=0,l_1}(\tau)}\text{ and }\nonumber \\
&\bra{\Psi_{n_2,l_2}(\tau)} z^*_{\tau'}\ket{\Psi_{n_1=0,l_1}(\tau)},
\end{align}
with $n_2\neq 0$. The expectation values involving $b_{\tau}$, $b^\dagger_{\tau}$ vanish because the Landau level index is different; the ones involving $a_{\tau}$ also vanish because the lowest Landau level is annihilated by it. Therefore, the only relevant contribution comes from the expectation value of $a^{\dagger}_{\tau}$, which will raise $n_1$ to $1$ and hence $n_2$ should be equal to $1$. Orthogonality implies $l_2=l_1$. Observe that only when $\tau'=\tau$ will $z_{\tau}$ have no $a^\dagger_{\tau}$. Furthermore, there is no choice of $\tau'$ that has $z^*_{\tau'}$ with no $a^\dagger_{\tau}$ dependence simply because that will imply transforming an annihilation operator into a creation operator ($a^{\dagger}_{\tau'}\sim a_{\tau}$) and vice-versa---such transformation cannot be unitary because there are no normalizable solutions for the equation $a^\dagger \ket{\Psi}=0$, for the standard harmonic oscillator problem. Let us then take $\tau'=\tau$. We find 
\begin{align}
&\bra{\Psi_{n_2,l_1}(\tau)} z_{\tau}\ket{\Psi_{n_1=0,l_1}(\tau)}=0 \text{ and }\nonumber \\
&\bra{\Psi_{n_2,l_1}(\tau)} z^*_{\tau}\ket{\Psi_{n_1=0,l_1}(\tau)}=\sqrt{2\tau_2}\ell_B \delta_{n_2,1}, \text{ with } n_2\neq 0.
\end{align}
Importantly, only in this case ($\tau'=\tau$) does $\Gamma^{\mathrm{int}}_{+}=0$  vanish exactly. Besides 
\begin{align}
\Gamma^{\mathrm{int}}_{-}=\frac{4\pi \mathcal{E}^2}{\hbar^2}\ell_B^2 l_F,    
\end{align}
such that, using the value of the area of the droplet $A= 2\pi \ell_B^2 l_F$, we get
\begin{align}
\frac{\Gamma^{\mathrm{int}}_+ -\Gamma^{\mathrm{int}}_-}{2A}= \frac{\mathcal{E}^2}{\hbar^2}c_1, \label{eq_quantized_CD}
\end{align}
with $c_1=-1$ the Chern number~\cite{tran:dauphi:grushin:zoller:goldman:2017,TranCooperGoldman18_PRA_circdichroLL}. We emphasize that this is a key result: the dichroic sum rule holds for arbitrary probe geometry $\tau'$, but the striking vanishing $\Gamma^{\mathrm{int}}_{+}=0$ --- the hallmark of perfect elliptic dichroism --- is achieved if and only if the probe is tuned to match the droplet's intrinsic complex structure, $\tau'=\tau$.

For completeness, in Appendix~\ref{sec: Gammapm for arbitrary tauprime}, we provide explicit expressions for $\Gamma^{\mathrm{int}}_{\pm}$ for arbitrary $\tau'$, showing this fact. Below we give an argument based on ideal Chern band theory which not only provides an alternative proof for the perfect elliptic dichroism result, but also generalizes it to arbitrary ideal Chern bands.
\subsection{Perfect elliptic dichroism and ideal Chern bands}
\label{subsec:perfect elliptic dichroism and ideal Chern bands}
In this Section, we extend the elliptic dichroic response from continuum Landau levels to general band insulators described by Bloch theory. For any filled set of bands, the integrated differential rate $\Gamma^{\mathrm{int}}_+ -\Gamma^{\mathrm{int}}_-$ is quantized according to the Chern number; see Ref.~\cite{tran:dauphi:grushin:zoller:goldman:2017} and Eq.~\eqref{eq_quantized_CD}. However, the stronger statement that one of the two (integrated) rates vanishes \emph{identically} is not generic: it is a special property that follows from the holomorphicity condition defining ideal Chern bands~\cite{wang:cano:millis:liu:yang:2021,mera:ozawa:2021,mera:ozawa:2021,wang:klevstov:liu:2023,junkai:wang:ledwith:vishwanath:parker:2023,mera:ozawa:2024}, and holds in particular when filling any set of $\mathcal{C}$ consecutive Landau levels $n=0, \dots, \mathcal{C}
-1$~\cite{ozawa:mera:2021,mera:ozawa:2021}. In ideal Chern bands, the complex structure determined by $\tau$ selects a single absorbing chirality, and the system exhibits perfect elliptic dichroism in the fullest sense: one rate is exactly zero, while the other is quantized by topology. The analysis below makes these statements precise.

One can describe the rates $\Gamma^{\mathrm{int}}_{\pm}$ using Bloch wavefunctions arising from magnetic translation symmetry. In the momentum space, the position operators become momentum derivatives. In particular
\begin{align}
z_{\tau}\longrightarrow i\left(-\frac{\partial}{\partial k_y} +\tau\frac{\partial}{\partial k_x} \right)=-2\tau_2 \frac{\partial}{\partial k^*_{\tau}},   
\end{align}
where $k_\tau= k_1 +\tau k_2$ is a complex variable in momentum space and $k^*_{\tau}$ its complex conjugate. Using this fact, we can write 
\begin{align}
\Gamma^{\mathrm{int}}_{+} \!&=\! \frac{8\pi \mathcal{E}^2 A \tau_2}{\hbar^2}\sum_{n \in \text{unocc.}}\!\sum_{m\in \text{occ.}} \int_{\BZ^2}\frac{d^2\bf{k}}{(2\pi)^2} \left|\bra{u_n}\frac{\partial}{\partial k^*_{\tau}}\ket{u_{m}} \right|^2 \nonumber \\
\Gamma^{\mathrm{int}}_{-} \!&\!=\frac{8\pi \mathcal{E}^2 A \tau_2}{\hbar^2}\sum_{n \in \text{unocc.}}\!\sum_{m\in \text{occ.}} \int_{\BZ^2}\frac{d^2\bf{k}}{(2\pi)^2} \left|\bra{u_n}\frac{\partial}{\partial k_{\tau}}\ket{u_{m}} \right|^2.
\end{align}
Introducing the orthogonal projectors $P=\sum_{m\in\text{occ.}}\ket{u_m}\bra{u_m}$ and $Q=1-P$, we can write
\begin{align}
&\Gamma^{\mathrm{int}}_{+} = \frac{8\pi \mathcal{E}^2 A \tau_2 }{\hbar^2} \int_{\BZ^2}\frac{d^2\bf{k}}{(2\pi)^2} \tr\left[\left(Q\frac{\partial P}{\partial k^*_{\tau}}P\right)^\dagger Q\frac{\partial P}{\partial k^*_{\tau}}P\right]  \text{ and }\nonumber \\
& \Gamma^{\mathrm{int}}_{-} = \frac{8\pi \mathcal{E}^2  A\tau_2}{\hbar^2}\int_{\BZ^2}\frac{d^2\bf{k}}{(2\pi)^2} \tr\left[\left(Q\frac{\partial P}{\partial k_{\tau}}P\right)^\dagger Q\frac{\partial P}{\partial k_{\tau}}P\right].
\label{eq: elliptic rates projector form}
\end{align}
Now use
\begin{align}
\frac{i}{2 \tau_2}dk_{\tau}\wedge dk^*_{\tau} = dk_x\wedge dk_y=d^2\bf{k},
\end{align}
so that after subtraction, we get
\begin{align}
&\frac{\Gamma^{\mathrm{int}}_{+} -\Gamma^{\mathrm{int}}_{-}}{2A}= \frac{\mathcal{E}^2}{\hbar^2} \frac{1}{2\pi} \nonumber \\
&\times\int_{\BZ^2} idk_{\tau}\wedge dk^*_{\tau}\!\left(\!\tr\left(\frac{\partial P}{\partial k_{\tau}} Q\frac{\partial P}{\partial k^*_{\tau}}\right)  \!-\!\tr\left(\frac{\partial P}{\partial k^*_{\tau}} Q\frac{\partial P}{\partial k_{\tau}}\right)\!\right)\nonumber \\
&=\frac{\mathcal{E}^2}{\hbar^2} \int_{\BZ^2} \tr\left(\frac{iF}{2\pi}\right)=\frac{\mathcal{E}^2}{\hbar^2} c_1(P).
\label{eq: elliptic dichroism band insulator}
\end{align}
Above, we identified the Berry curvature 
\begin{align}
F&=\tr(dPQ\wedge dP)\nonumber \\
&=\left[\tr\left(\frac{\partial P}{\partial k_{\tau}}Q\frac{\partial P}{\partial k^{*}_{\tau}}\right)-\tr\left(\frac{\partial P}{\partial k^*_{\tau}}Q\frac{\partial P}{\partial k_{\tau}}\right)\right]dk_{\tau}\wedge dk_{\tau}^*,
\end{align}
and $c_1(P)=\int_{\BZ^2} \tr\left(\frac{iF}{2\pi}\right)$ is the first Chern number of the occupied band, described by $P$.

As discussed in Refs.~\cite{ozawa:mera:2021, wang:cano:millis:liu:yang:2021,mera:ozawa:2021,mera:ozawa:2024,wang:klevstov:liu:2023,liu:mera:fujimoto:ozawa:wang:2025}, the anisotropic lowest Landau level is an ideal Chern band. This fact is equivalent to the following condition
\begin{align}
Q\frac{\partial P}{\partial k^*_{\tau}} P=0, 
\label{eq: ideal Chern band condition}
\end{align}
which follows from the saturation of the trace condition. It also means that we can find, locally, a basis of Bloch wavefunctions for the band determined by $P$ which is holomorphic in $k_{\tau}$.
This relation is also equivalent to the vortexability condition~\cite{ledwith:vishwanath:parker:2023,fujimoto:parker:dong:khalaf:vishwanath:ledwith:2025}
\begin{align}
Q[z_{\tau},P]P=0 \iff z_{\tau} P=Pz_{\tau} P,
\end{align}
which essentially indicates that $z_{\tau}$ preserves the lowest Landau level. 

This brings us to the central result of this Section:~the ideal Chern band condition Eq.~\eqref{eq: ideal Chern band condition} directly and immediately implies the exact vanishing $\Gamma^{\mathrm{int}}_{+}=0$, since $\tr\left[\left(Q\frac{\partial P}{\partial k^*_{\tau}}P\right)^\dagger Q\frac{\partial P}{\partial k^*_{\tau}}P\right]=0$: the holomorphicity of the occupied Bloch states with respect to $k_{\tau}$ is precisely the condition that kills one chiral absorption rate entirely. A version of this statement was also established for the isotropic case, i.e. $\tau=i$, in Ref.~\cite{junkai:wang:ledwith:vishwanath:parker:2023} and referred to as perfect circular dichroism. It is then natural to refer to the current phenomenon of $\Gamma^{\mathrm{int}}_{+}=0$ as \emph{perfect elliptic dichroism}.

\subsection{Flux-space representation for many-body systems}
We note that the proof of perfect elliptic dichroism based on the projector form of the rates in Eq.~\eqref{eq: elliptic rates projector form} and on the holomorphicity condition Eq.~\eqref{eq: ideal Chern band condition} can also be adapted to the case of many-body states, in particular for fractional quantum Hall states, such as Laughlin states, where the center-of-mass coordinates decouple from the relative coordinates.
To do so, following Ref.~\cite{goldman2024relating}, one works on a finite system with twisted periodic boundary conditions, described by twist-angles $\theta_a$. Equivalently, we can work with periodic boundary conditions, absorbing the twist-angles into the vector potential $eA_a\to eA_a -\hbar\theta_a/L_a$, where $L_a$ is the size of the system in the $a$th direction. The effect of turning on the periodic electric field can be understood, in an appropriate gauge, in terms of periodically modulating the twist-angles $E^{\pm}_a(t)=-\frac{\hbar}{e}\partial_t \theta_a(t)/L_a$. We set $L_1=L_2=L$. We have
\begin{align}
\delta H(\theta_1,\theta_2)=\pm \frac{2\mathcal{E}L\sqrt{\tau_2}}{\hbar\omega} \left(\frac{\partial H}{\partial \theta_{\tau}^*} e^{\mp i\omega t} +\text{H.c.}\right),   
\end{align}
which yields
\begin{align}
\Gamma^{\mathrm{int}}_{+} &=\frac{8\pi L^2 \mathcal{E}^2}{\hbar^2}\frac{\tau_2}{\mathrm{rank}(P)}\tr\left(\frac{\partial P}{\partial \theta_{\tau}}Q\frac{\partial P}{\partial \theta_{\tau}^*}\right),\nonumber \\
\Gamma^{\mathrm{int}}_{-} &=\frac{8\pi L^2 \mathcal{E}^2}{\hbar^2}\frac{\tau_2}{\mathrm{rank}(P)}\tr\left(\frac{\partial P}{\partial \theta_{\tau}^*}Q\frac{\partial P}{\partial \theta_{\tau}}\right),
\label{eq: integrated rates twist-angle}
\end{align}
where $\theta_{\tau}=\theta_1+\tau\theta_2$ is the complex coordinate in twist-angle space, $P=\sum_{\alpha=1}^{\mathrm{rank}(P)}\ket{\Psi_{\alpha}}\bra{\Psi_{\alpha}}$ denotes the projection onto the subspace of Laughlin states consistent with the boundary condition prescribed by $\theta_a$ and $\{\ket{\Psi_{\alpha}}\}$ denotes an orthonormal basis of Laughlin states. Here $\mathrm{rank}(P)=\tr(P)=\frac{1}{\nu}$, where $\nu$ is the filling fraction, is the degeneracy of this subspace. Its appearance can be understood in two ways. One way is to consider each ground state at a time and then perform an average~\footnote{Note however that the individual $\ket{\Psi_{\alpha}}$'s do not determine line bundles on their own, since besides a phase factor there will be a unitary monodromy matrix, as each $\theta_a\to \theta_a+2\pi$. This monodromy turns out to be $\theta_a$-independent, which allows the Berry connection determined by $P$ to be projectively flat and produces the correct result.}. Alternatively, we can see this as a consequence of the density matrix of the system being described by $P/\tr(P)$. This latter object is well-defined and gauge-invariant. In the thermodynamic limit, since the Berry curvature in twist-angle space is uniform,
\begin{align}
F &=\tr\left(dPQ\wedge dP\right)\\
&=\left[\tr\left(\frac{\partial P}{\partial \theta_{\tau}}Q\frac{\partial P}{\partial \theta^{*}_{\tau}}\right)-\tr\left(\frac{\partial P}{\partial \theta^*_{\tau}}Q\frac{\partial P}{\partial \theta_{\tau}}\right)\right]d\theta_{\tau}\wedge d\theta_{\tau}^*\nonumber \\
 &=\frac{c_1(P)}{4\pi\tau_2}d\theta_{\tau}\wedge d\theta^*_{\tau},  
\end{align}
where $c_1(P)=-1$ is the first Chern number of the Laughlin vector
bundle over the twist-angle space. Thus, using Eq.~\eqref{eq: integrated rates twist-angle}, 
\begin{align}
\frac{\Gamma^{\mathrm{int}}_{+}-\Gamma^{\mathrm{int}}_{-}}{2 A}=\frac{\mathcal{E}^2}{\hbar^2}\mathcal{C}_{\mathrm{MB}},
\end{align}
where $A=L^2$ is the area of the sample, and $\mathcal{C}_{\mathrm{MB}}$ denotes the many-body Chern number~\cite{goldman2024relating,repellin2019detecting,unal_2025}.
Here, we used that
\begin{align}
\mathcal{C}_{\mathrm{MB}}=\frac{c_1(P)}{\mathrm{rank}(P)}=-\nu,
\end{align}
where $\mathrm{rank}(P)=\tr(P)=\nu^{-1}$ is the rank of the Laughlin vector bundle. Because the Laughlin wavefunctions are holomorphic in the complex variable $\theta_{\tau}=\theta_1+\tau \theta_2$ (they are essentially Jacobi theta functions at level $\nu^{-1}$, describing the center of mass part of the wavefunction~\cite{haldane:rezayi:85,klevtsov:2016}), it follows that $Q\frac{\partial P}{\partial \theta_{\tau}^{*}}=0$, so they describe an ideal band in twist-angle space. Hence $\Gamma^{\mathrm{int}}_{+}=0$, as before, and we have perfect elliptic dichroism. 

This proof also reproduces the result in Eq.~\eqref{eq: elliptic dichroism band insulator} if one considers $P$ to be the many-body projector corresponding to filling a given band. In this case $\mathrm{rank}(P)=1$, so $\mathcal{C}_{\mathrm{MB}}$ is an integer and coincides with the Chern number from ordinary band theory. The single-particle ideal band condition in Eq.~\eqref{eq: ideal Chern band condition} is seen to imply $Q\frac{\partial P}{\partial \theta_{\tau}^{*}}=0$ for the many body projection—because the twisted boundary condition appears as a shift in each of the individual Bloch momenta and holomorphicity in $k_{\tau}$ implies holomorphicity in $\theta_{\tau}$—and perfect elliptic dichroism then follows.

We emphasize that the perfect elliptic dichroism discussed above is not a statement about the intrinsic guiding-center geometry of the correlated liquid. In the continuum quantum Hall problem, the latter is encoded in the guiding-center degrees of freedom and, for an interacting state, may result from a nontrivial compromise between the Landau-orbit metric and the metric selected by the interaction. By contrast, the vanishing of one of the elliptic absorption rates follows from the holomorphic dependence of the relevant projector on the complex twist-angle variable $\theta_\tau=\theta_1+\tau\theta_2$. In this sense, the result is insensitive to the emergent metric of the correlated fluid, whenever the many-body ground-state projector satisfies the corresponding holomorphicity condition in flux space. This distinction is particularly transparent for Laughlin states, where the center-of-mass sector is described by theta functions (see Refs.~\cite{haldane:rezayi:85,klevtsov:2016}) holomorphic in $\theta_\tau$, leading to $Q\frac{\partial P}{\partial \theta^*_\tau}P=0$ and hence to $\Gamma^{\mathrm{int}}_+=0$.
Let us also stress that Galilean invariance is not essential for this mechanism. In a Galilean-invariant Landau-level problem, the cyclotron dynamics is governed by the quadratic form $g^{ab}\pi_a\pi_b$, where $g^{ab}$ is the inverse cyclotron metric. This metric determines a distinguished complex coordinate, and hence a complex structure, associated with the cyclotron motion, thereby fixing, up to an overall phase, the Landau-level ladder operators and the elliptic probe naturally adapted to the lowest Landau level. 
The same complex structure also underlies the ideal-band description obtained by filling the first $r$ Landau levels, as well as the holomorphic flux-space description of Laughlin states. However, the argument above only requires the existence of a complex structure with respect to which the occupied projector is holomorphic. Therefore the same reasoning applies to ideal Chern bands, including lattice systems without continuous translation or rotation symmetry, provided they satisfy the ideal-band/vortexability condition $Q\frac{\partial P}{\partial k^*_\tau}P=0$. In such cases the complex coordinate $k_\tau$, or equivalently the associated real-space coordinate $z_\tau$, selects an emergent conformal class of metric adapted to the band geometry, but this should not be confused with microscopic Galilean invariance.
\section{Application to the Harper-Hofstadter Model}
\label{sec:application}
As an application, we now consider the problem of free-fermions in a square Harper-Hofstadter lattice governed by the tight-binding Hamiltonian
\begin{equation}
	\label{eq:hofst_latt}
    \begin{split}
    H_{\rm HH} =& -J_x \sum_{m,n} c_{m+1,n}^\dagger c_{m,n} \\& -J_y \sum_{m,n} e^{i2\pi \alpha m}c_{m,n+1}^\dagger c_{m,n}+{\rm h.c.}
    \end{split}
\end{equation}
Here, $c_{m,n}^{(\dagger)}$ are the annihilation (creation) operators of a single fermion at the given lattice site $(m,n)$ and $2\pi \alpha$ represents the magnetic flux per plaquette.
We plot the density of the system's ground state in Fig.~\ref{fig:dichroism}(a); it can be seen that the bulk's density, $\langle{n_{m,n}\rangle}=\langle{c^\dagger_{m,n}c_{m,n}\rangle}$ reaches the homogeneous expected value $n_0=\frac{a_0^2}{2\pi\ell_B^2}$ in the bulk; here $\ell_B=a_0/\sqrt{2\pi\alpha}$ is the magnetic length, and $a_0$ the lattice spacing. Conversely, it can be noted that the density near the boundaries reflects the hopping anisotropy $J_x\neq J_y$.

This lattice model can be mapped onto an effective continuum description when the magnetic length $\ell_B$ 
is much larger than the lattice spacing $a_0$, provided we focus on its low-energy, long-wavelength features.
Strictly speaking, a direct operator equality between the lattice Hamiltonian $H_{\rm HH}$ and the one in the continuum, $H$, is precluded by a fundamental mismatch of their underlying Hilbert spaces; this is reconciled by understanding the correspondence as an asymptotic operator equivalence. 
In particular, as we discuss in further detail in Appendix~\ref{eq:continuum limit}, at low-energies and long-wavelengths the eigenstates $\ket{\psi_i}=\sum_{\bf R} C_{{\bf R},i}\ket{{\bf R}}$ of the Harper-Hofstadter model $H_{\rm HH}$ will feature probability amplitudes $C_{{\bf R},i}$ for the given lattice site ${\bf R}=(m,n)$ that vary smoothly over the lattice scale $a_0$; meaning, these discrete amplitudes can be coarse-grained to smooth ones, $C_i({\bf R})$, the latter being in fact eigenvectors of an anisotropic Landau Hamiltonian Eq.~\eqref{eq: anisotropic LL Hamiltonian},
\begin{equation}
    \label{eq:ll_limit}
    H = \frac{1}{2\meff} g^{ab}\pi_a \pi_b + \delta H,
\end{equation}
up to an irrelevant energy offset which we here dropped. The effective mass parameter is here set by $\meff^{-1}=2a_0^2\sqrt{J_xJ_y}/\hbar^2$. 
The spatial anisotropy which is manifest in the hopping amplitudes of $H_{\rm HH}$, gives rise to a diagonal unimodular Landau-orbit metric
$(g^{ab})=\mathrm{diag}(\beta,\beta^{-1})$, where we introduced the parameter $\beta=\sqrt{J_x/J_y}$; the Hamiltonian Eq.~\eqref{eq:ll_limit} with the metric $g_{ab}$ corresponds to choosing the value of the modular parameter to be 
\begin{equation}
    \label{eq:hh_metric}
    \tau= i \beta^{-1}= i \sqrt{J_y/J_x}.
\end{equation}
Notice how the Landau Hamiltonian (the first term on the left-hand side of Eq.~\eqref{eq:ll_limit}) trivially commutes with the generator of rotations $L=\frac{1}{2}g^{ab}\pi_a  \pi_b$~\cite{Park_2014}. 

Such approximation is useful since it bridges 
the lattice setup considered here directly with the continuum geometric framework developed in Sec.~\ref{sec:the metric of quantum hall settings} and ~\ref{sec: elliptic dichroic response}, allowing us to benchmark numerically exact results obtained from the exact diagonalization of the lattice system --- which are highly relevant for ongoing experiments with both cold-atom and photonic platforms~\cite{cooper:dalibard:spielman:2019,ozawa:price:amo:goldman:etal:2019} --- 
against the analytical theory. 
Furthermore, the term $\delta H$, which explicitly accounts for the high-energy corrections stemming from the discreteness of the underlying grid, allows for the (perturbative) analysis of the effects of the lattice; indeed,
when the lattice spacing is much smaller than the magnetic length the anharmonic corrections $\delta H$ can be organized as an ordered perturbative expansion in $a_0/\ell_B$ (see Appendix~\ref{eq:continuum limit});
\begin{subequations}
\label{eq:lattice hamiltonian perturbation}
in particular, at order $\mathcal O(a_0^4/\ell_B^4)$ (notice how the kinetic momenta scale as $\ell_B^{-1}$, see the commutator Eq.~\eqref{eq:pis_commutator}) one has
\begin{equation}
    \label{eq:perturbation lowest_order}
    \begin{split}
    \delta H_1&=-\frac{a_0^4}{12\hbar^4} \left(J_x  \pi_x^4+J_y  \pi_y^4\right) = \\&=-\frac{a_0^2 }{24\hbar^2m_{\rm eff}} \left( \beta \pi_x^4+ \beta^{-1} \pi_y^4\right),
    \end{split}
\end{equation}
while at the next one $\mathcal O(a_0^6/\ell_B^6)$,
\begin{equation}
    \label{eq:perturbation first_order}
	\delta H_2 = \frac{a_0^4}{720\hbar^4m_{\rm eff}} \left( \beta \pi_x^6 + \beta^{-1} \pi_y^6\right).
\end{equation}    
\end{subequations}
Notice crucially how these corrections do not commute with $L$, which would be the case if the kinetic energy was a function of $g^{ab}\pi_a\pi_b$ alone. 
As a consequence the system can be described by a unique metric only at lowest order in $a_0/\ell_B$ expansion, when the Chern bands of $H_{\rm HH}$ exactly become Landau levels. In other words, the Harper-Hofstadter Hamiltonian cannot be written as a function of $L$ alone:
a ``tension'' emerges at finite non-zero $a_0/\ell_B$. For example, neglecting the $\mathcal O(a_0^6/\ell_B^6)$ contribution coming form $\delta H_2$ for simplicity, the low-energy effective Hamiltonian $H=H_0+\delta H_1$ can be exactly rewritten as $H=f_0(g_0^{ab}\pi_a\pi_b)+f_1(g_1^{ab}\pi_a\pi_b)$ with
\begin{align}
    f_0(x)&=x/2\meff - \lambda x^2\\
    f_1(x)&=-\lambda x^2,
\end{align}
where the parameter $\lambda$ is fixed by $\lambda^{-1}=24\hbar^2(\beta+\beta^{-1})\meff/a_0^2$.
Notice crucially how besides the inverse metric $(g_0^{ab})={\rm diag}(\beta,\beta^{-1})$, a new quantity $(g_1^{ab})={\rm diag}(1,-1)\neq (g_0^{ab})$ naturally appears.
This $g_1$ should not be understood as defining a second metric by itself, but rather as the leading-order perturbation $\delta g$ of the continuum metric $g_0$. Indeed, we will see later that the same tensor appears at lowest order in $\alpha$ when computing perturbatively the expectation value $\frac{1}{2}\langle \pi_a\pi_b+\pi_b\pi_a\rangle$,
see Eq.~\eqref{eq: perturbed Landau-orbit metric}. One may therefore expect a new ``effective'' metric --- the Landau-orbit metric --- to emerge in the ground-state wavefunction~\cite{Haldane_2011}, as a compromise between the continuum Landau-level metric and the finite-lattice-spacing correction. This emergent metric should then control the geometric response properties of the system.
In the next sections, we show how lattice effects modify the perfect elliptic dichroic response expected from the continuum-limit analysis, and how these modifications can indeed be interpreted as arising from a renormalized Landau-orbit metric.

\subsection{Elliptic dichroism in the Harper-Hofstadter model}
We subject the system to a time-dependent dichroic probe Eq.~\eqref{eq: dichroic probe}
\begin{equation}
    \label{eq:hh_probe}
	V_{\pm}(r) = - \Re\left[2\mathcal{E} e^{-i \omega t} \left( e^{-\frac r 2} x \pm i e^{\frac r 2} y \right)\right],
\end{equation}
where the  amplitude $2\mathcal{E}$ and the squeezing-parameter $r$ are both real. This can indeed be conveniently recast in the form of Eq.~\eqref{eq: dichroic probe},
\begin{align}
V_{\pm}(r) &= - \Re\left[2\mathcal{E} e^{-i \omega t} \left( e^{-\frac r 2} x \pm i e^{\frac r 2} y \right)\right] \nonumber \\
&=\pm\left[-2\mathcal{E} e^{\frac{r}{2}}\;\mathrm{Re}\left[e^{\mp i(\omega t +\frac{\pi}{2})}(-y+ie^{-r}x)\right]\right]\nonumber\\
&=\pm \left[-\frac{2\mathcal{E}}{\sqrt{\tau'_2}}  \; \mathrm{Re}\left(z_{\tau'} e^{\mp i(\omega t+\delta)}\right)\right],
\end{align}
with $z_{\tau'}=-y+\tau' x$, $\tau'=i e^{-r}$, $\tau_2'=e^{-r}$ and $\delta=\pi/2$. Since the overall sign of $V_{\pm}$ and the phase shift do not affect the dichroic rates according to Fermi's golden rule, we can compare with the results obtained in Sec.~\ref{sec: elliptic dichroic response}.

According to the above discussion (cf. Eqs. ~\eqref{eq: dichroic probe} and ~\eqref{eq:probe_decomposition}), we can split such a probe into its kinetic-momenta and guiding-center components using $R^a-\epsilon^{ab}\pi_b/B=r^a$, $(r^a)=(x,y)$ (see Eq.~\eqref{eq:guiding_centers}):
\begin{equation}
    V_{\pm}(r) = V_{\pm}^{\rm km}(r) + V_{\pm}^{\rm gc}(r),
\end{equation}
where
\begin{align}
    \label{eq:probe_km}
    V_{\pm}^{\rm km}(r) &= 	- \Re\left[\frac{2\mathcal{E}\ell_B^2}{\hbar} e^{-i \omega t} \left( e^{-\frac r 2} \pi_2 \mp i e^{\frac r 2} \pi_1 \right)\right],
    \\
    \label{eq:probe_gc}
    V_{\pm}^{\rm gc}(r) &= 	- \Re\left[2\mathcal{E} e^{-i \omega t} \left( e^{-\frac r 2} R^1 \pm i e^{\frac r 2} R^2 \right)\right].
\end{align}
Note that 
 $V_{\pm}^{\rm gc}(r)$ contributes to the excitation rates through transitions localized at the edge of the system: indeed, in the notation of Eq.~\eqref{eq:fock_states}, the $\Delta n=0$, $\Delta l=\pm 1$ transitions induced by the probe can only occur close to the Fermi point $l_F$, i.e. at the edge of the system. On the other hand, those coming from $V_{\pm}^{\rm km}(r)$ (which instead obey $\Delta n=\pm1$, $\Delta l=0$ selection rules) are genuinely bulk, since every particle can be promoted to the higher Landau level. In the following, and as in Sec.~\ref{sec: elliptic dichroic response},  we disregard the edge contributions and focus on the genuine bulk response.

While in general both chiral rates associated with the $\pm$ polarizations of the elliptic probe Eq.~\eqref{eq:probe_km} are non-vanishing, the discussion above established that a special situation arises when the probe metric is tuned to match the intrinsic metric of the droplet: at a specific value of the probe ellipticity $r$, the absorption rate associated with the $V_+^{\rm km}$ polarization vanishes exactly, signaling the onset of perfect elliptic dichroism. To see this explicitly, it is instructive to rewrite the kinetic-momenta part of the probe, Eq.~\eqref{eq:probe_km}, directly in terms of the bosonic ladder operators $a_\tau$ defined in Eq.~\eqref{eq: complex structure tau orbital operators}:
\begin{equation}
	\begin{split}
		V_{\pm}^{\rm km}(r) =
		\Re\Biggl[\mp i\sqrt{2}\mathcal{E}\ell_B e^{-i \omega t}\biggl[ &\left( e^{\frac{r-r_{0}} 2} \pm e^{-\frac{r-r_{0}}{2}} \right) a_\tau +\\+&\left( e^{\frac{r-r_{0}} 2} \mp e^{-\frac{r-r_{0}}{2}} \right) a^\dagger_\tau \biggr] \Biggr],
	\end{split}
\end{equation}
where $r_0=\log(\beta)$ is the ``intrinsic-metric" squeezing parameter (cf. Eq.~\eqref{eq:hh_metric}). From this expression it is clear that the ``perfect dichroic" condition is met when the externally tunable parameter $r$ obeys
\begin{equation}
    \label{eq:metric_matching}
    r=r_0=\log(\beta),
\end{equation}
since, as we discussed in Sec.~\ref{subsec:quantized circular dichroic response}, $V_+^{\rm km}(r=r_0)\propto a_\tau$ in this limit.

\begin{figure}[t]
  \includegraphics[width=1\linewidth]{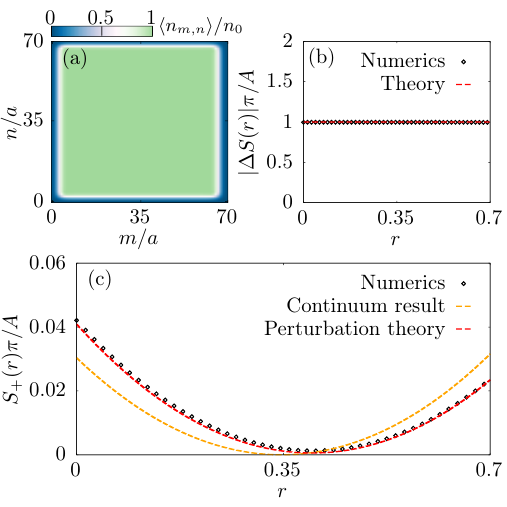}
    \caption{
        {(a) Density $\Braket{\hat n_{m,n}}$ of a system of $N_F=333$ fermions occupying the lowest band of the Harper-Hofstadter model on the square $70\times70$ lattice Eq.~\eqref{eq:hofst_latt} with anisotropic hopping frequencies $J_x/J_y=2$ and $\alpha=2/25$. 
        The density has been normalized to the bulk value $n_0=1/2\pi\ell_B^2$.
        (b) As a function of the probe ellipticity parameter $r$ (cf. Eq.~\eqref{eq:hh_probe}), we plot the numerically computed dichroic rate difference $\Delta S=S_+-S_-$ (black circles), comparing it to the expected value Eq.~\eqref{eq:dichro_sum_rule} (red dashes). To isolate the bulk value, we focused only on high-energy transitions, filtering away the edge low-energy contribution~\cite{unal_2025}.
        (c) As a function of $r$, we plot $S_+$ comparing it with the theoretical (perturbative) prediction  Eq.~\eqref{eq:rates_pt_Splus}.
        }
}
    \label{fig:dichroism}
\end{figure}

While the bulk response is dominated by the kinetic-momenta component Eq.~\eqref{eq:probe_km}, in our numerical study we look at the full frequency-integrated excitation rates associated with the coordinate-based dichroic probe Eq.~\eqref{eq:hh_probe}. Within Fermi's golden rule (cf.\ Eq.~\eqref{eq: dichroic rates}), these read
\begin{subequations}
\begin{equation}
    \label{eq:rates_spm}
    \Gamma^{\mathrm{int}}_\pm = \frac{2\pi \mathcal{E}^2}{\hbar^2} S_\pm
\end{equation}    
where we introduced
\begin{equation}
    S_\pm \equiv \sum_{n\neq0} \left|\int d^2\boldsymbol{r} (e^{-\frac r2}x\pm i e^{\frac r2}y)\bra{n}n(\boldsymbol{r})\ket{0}\right|^2.
\end{equation}
\end{subequations}
Here, $n({\bf r})=\sum_{i=1}^N\delta({\bf r}-{\bf r}_i)$ is the density operator. 
The integrated rate difference $\Delta\Gamma=\Gamma_+^{\rm int}-\Gamma_-^{\rm int}$ is associated to the quantized dichroism [Eq.~\eqref{eq_quantized_CD}] discussed in Sec.~\ref{subsec:quantized circular dichroic response}. As a consequence, 
\begin{equation}
\Delta S=S_+-S_- = \frac{A}{\pi} c_1.\label{eq_quantized_CD_bis}
\end{equation}

As previously discussed, the bulk response is dominated by the kinetic-momenta component Eq.~\eqref{eq:probe_km} of the probe Eq.~\eqref{eq:hh_probe}: under such approximation, 
these rates can then be analytically computed. With standard second order perturbation theory in $\delta H$ (defined by Eq.~\eqref{eq:lattice hamiltonian perturbation}; notice also how the second order corrections in $\delta H_1$ scale in the same way as first order corrections in $\delta H_2$) 
we obtain, for the cyclotron-only part (i.e. neglecting edge contributions),
\begin{subequations}
    \label{eq:rates_pt}
    \begin{equation}
	\label{eq:rates_pt_Splus}
	S_+ =\frac{A}{\pi} \Biggl( \sinh^2\left(\frac{r-r_0}{2}\right) + \delta S(r)\Biggr)
	\end{equation}
    \begin{equation}
    \label{eq:rates_pt_Sminus}
	S_- =\frac{A}{\pi}\Biggl( \cosh^2\left(\frac{r-r_0}{2}\right) + \delta S(r)\Biggr),
    \end{equation}
\end{subequations}
where $\delta S=\delta S_1+\delta S_2$ is given by   
\begin{subequations}
\begin{equation}
	\begin{split}
	\delta S_1(r) =& \alpha A_0 \sinh(r-r_0)
	\end{split}
\end{equation}
and
\begin{equation}
	\begin{split}
	\delta S_2(r) =
	\alpha^2\Biggl[A_1&\cosh(r-r_0) + A_2 \sinh(r-r_0)\Biggr],
	\end{split}
\end{equation}
\end{subequations}
and for convenience we introduced $A_0=\frac{\pi}{8}(\beta^{-1}-\beta)$, $A_1=\frac{\pi^2}{384}(7\beta^2-10+7\beta^{-2})$ and $A_2=\frac{\pi^2}{24}(\beta^{-2}-\beta^2)$.

As anticipated above, we find that $\Delta S = S_+-S_-$ is conserved, since the correction terms $\delta S(r)$ cancel out:
\begin{equation}
    \label{eq:dichro_sum_rule}
    \Delta S = -\frac{A}{\pi}.
\end{equation}
This result is in accordance with the general argument exposed in Sec.~\ref{sec: elliptic dichroic response}:~no matter the probe's metric (here parametrized by the squeezing parameter $r$), the integrated differential rate is quantized according to the many-body Chern number, here with $c_1=-1$; see Eq.~\eqref{eq_quantized_CD_bis}.

We demonstrate this result in Fig.~\ref{fig:dichroism}(b), where we find that the absolute value of the dichroic rate difference, $|\Delta S(r)|$, is very close to the expected value $A/\pi$, for every value of the probe ellipticity $r$.

Then, we verify that $S_+(r=r_0)$ is non vanishing, but indeed acquires a finite value $\propto\alpha^2$  owing to lattice corrections which forbid a unique metric to be introduced at all energy-scales.
Notice furthermore how the minimum of $S_+(r)$ gets horizontally shifted from $r_0$. 
This is numerically demonstrated in Fig.~\ref{fig:dichroism}(c); in particular, we compare the numerically obtained $S_+(r)$ (black circles) with the perturbative expansion of $S_+(r)$ of Eq.~\eqref{eq:rates_pt_Splus} (red dashes) and with its zero-order, continuum-limit result $S_+(r)=\frac{A}{\pi} \sinh^2((r-r_0)/2)$ (orange dashes) (cf. Eq.~\eqref{eq:rates_pt_Splus}); It can clearly be seen that the former expression correctly accounts for the deviation of the numerics from the continuum limit.

\begin{figure}[t]
  \includegraphics[width=1\linewidth]{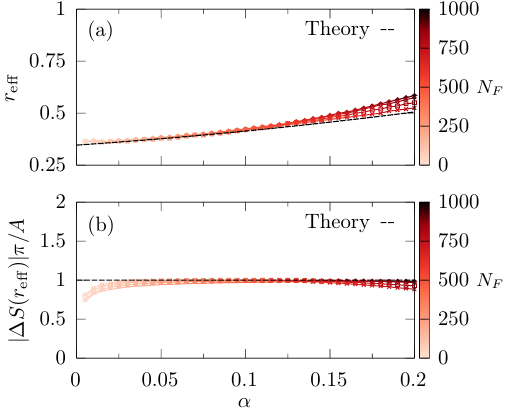}
    \caption{
        {(a) As a function of the flux per plaquette $\alpha$, we show the numerically computed position of the minimum of $S_+(r)$ is compared with the theoretical prediction Eq.~\eqref{eq:effective_dr}, for various filling fractions $N_F/\mathcal N$ 
        of the lowest band of the $70\times70$ Harper-Hofstadter lattice Eq.~\eqref{eq:hofst_latt} with anisotropic hopping rates $J_x/J_y=2$. Here, $\mathcal N=\alpha N_xN_y$ is the Landau level degeneracy and we consider
        $N_F=0.80 \mathcal N$ (crosses), $N_F=0.85 \mathcal N$ (squares), $N_F=0.90 \mathcal N$ (circles), $N_F=0.95 \mathcal N$ (triangles).
        In (b) we check the dichroic sum rule Eq.~\eqref{eq:dichro_sum_rule} at the position of the minimum.
        }
}
    \label{fig:dichroism_sweep}
\end{figure}

In particular, according to the second-order perturbative expansion above, the new position of the minimum of $S_+(r)$, $r_{\rm eff}$, becomes
\begin{equation}
    \label{eq:effective_dr}
    \begin{split}
    r_{0}' - r_0 = \delta r =& {\rm arctanh}\left(-\frac{ \alpha A_0+\alpha^2 A_2}{\frac{1}{2}+\alpha^2 A_1}\right)=\\=&
    -2A_0\,\alpha-2A_2\,\alpha^2 + \mathcal{O}(\alpha^3):
    \end{split}
\end{equation}
this is a “renormalized” ellipticity that the probe needs to have in order to minimize the response from the $V_+(r_{\rm eff})$ polarization of the probe Eq.~\eqref{eq:hh_probe} and achieve the “perfect” dichroic response.
We demonstrate the predicted magnetic field dependence of $r_{\rm eff}$ in Fig.~\ref{fig:dichroism_sweep}(a), where we compare the numerical results for the minimum of $S_+(r)$, $r_{\rm eff}$ (points) with the perturbative prediction Eq.~\eqref{eq:effective_dr} (black dashes)
as a function of the flux per plaquette $\alpha$ for various fillings of the lowest band.
As expected, the prediction works especially good at small $\alpha$, since $a_0/\ell_B$ is then small. The deviations observed at the smallest values analyzed should be traced to the fact that the number of particles occupying the fixed-size lattice dramatically decreases as $\alpha\rightarrow0$.
The same effect can be seen in Fig.~\ref{fig:dichroism_sweep}(b), where we benchmark the sum-rule Eq.~\eqref{eq:dichro_sum_rule} at the numerically obtained value $r=r_{\rm eff}$.

For gaining some insight later, it is useful to approximate Eq.~\eqref{eq:effective_dr} to lowest order in $\alpha$:
\begin{equation}
    \label{eq:effective_dr_linear}
    \delta r = \frac{\pi\alpha}{4}\left(\beta -\beta^{-1}\right).
\end{equation}

\subsection{The emergent metric}
We now show how the shift $r_{0}'=r_0+\delta r$ can be interpreted as the emergence of a new effective metric; the fact that $S_+$ has its minimum at this value can then be interpreted as the probe's metric matching the emergent metric.

Notice first how, as a consequence of the structure of the Landau Hamiltonian Eq.~\eqref{eq: anisotropic LL Hamiltonian}, the metric naturally emerges from the variances of the kinetic momenta, as $\frac{1}{2}\bra{0}\pi_a\pi_b+\pi_b\pi_a\ket{0}=g_{ab} \frac{\hbar^2}{2\ell_B^2}$. Explicitly,
\begin{align}
	\label{eq:metric_from_variance}
		&\bra{0}\pi_x^2\ket{0} = \beta^{-1}\,\frac{\hbar^2}{2\ell_B^2}
		\\
		&\bra{0}\pi_x\pi_y + \pi_y\pi_x\ket{0} = 0
		\\
		&\bra{0}\pi_y^2\ket{0} = \beta\,\frac{\hbar^2}{2\ell_B^2}.
\end{align}
Here $\ket{0}$ is a shorthand label for a generic lowest Landau level eigenfunction.
At first perturbative order in $\delta H$ (thus accounting for the aforementioned lattice corrections) we get
\begin{align}
		&\bra{0'}\pi_x^2\ket{0'} = \frac{\hbar^2}{2\ell_B^2} \,\,\beta^{-1}\left[1 -\frac{2\pi\alpha}{8} \left( \beta -\beta^{-1} \right) \right],
		\nonumber\\
		&\bra{0'}\pi_x\pi_y+\pi_y\pi_x\ket{0'} = 0,\nonumber
		\\
		&\bra{0'}\pi_y^2\ket{0'} = \frac{\hbar^2}{2\ell_B^2} \,\, \beta \left[ 1+\frac{2\pi\alpha}{8} \left( \beta -\beta^{-1}\right) \right].
        \label{eq: perturbed Landau-orbit metric}
\end{align}
Here $\ket{0'}$ is a shorthand label for a generic dressed lowest Landau level eigenfunction.
Notice how the third one is the (perturbative) reciprocal of the first one; upon comparing these equations with the unperturbed ones Eq.~\eqref{eq:metric_from_variance} we are led to the introduction of a renormalized $\beta$
\begin{equation}
	\label{eq:renormalized_metric}
	\beta_{\rm eff} = \beta \left[ 1+\frac{2\pi\alpha}{8} \left( \beta  -\beta^{-1} \right) \right] ,
\end{equation}
in terms of which the (inverse) effective metric (at lowest order in $\alpha$) reads
\begin{equation}
	(\widetilde g_{\rm eff}^{ab})=\begin{pmatrix} \beta_{\rm eff} & 0\\   0 & \beta^{-1}_{\rm eff}\end{pmatrix},
\end{equation}
which effectively takes into account the (quartic) lattice corrections to the Landau Hamiltonian. 
Notice how this effective metric remains diagonal and depends on the magnetic field, $\alpha=\frac{1}{2\pi}(a_0/\ell_B)^2$.

Notice that the minimum of the ellitpic dichroic rate $S_+$ (i.e. the ``perfect dichroism") is, as anticipated, set by this effective metric (at this order in the magnetic field): since $r_{\rm eff} = \log(\beta_{\rm eff})$ (cf. Eq.~\eqref{eq:metric_matching}), at lowest order in $\alpha$ we recover
\begin{equation}
	\begin{split}
	r_{\rm eff} & = r_0 + \log\left[ 1+\frac{2\pi\alpha}{8} \left( \beta -\beta^{-1} \right)\right] 
	\simeq\\
	&\simeq 
	r_0 +\frac{2\pi\alpha}{8} \left( \beta  -\beta^{-1}\right) = r_0+\delta r.
	\end{split}
\end{equation}

\section{Conclusions and outlook}
\label{sec:outlook}

In this work, we introduced the notion of perfect elliptic dichroism as a direct probe of the geometric structure of quantum Hall systems. By tailoring the geometry of the driving field and matching it to the intrinsic complex structure of the quantum Hall droplet, we identified a regime where the response becomes maximally asymmetric: one chiral component of the elliptic drive is completely inactive, while the other exhibits a quantized response fixed by the Chern number. This striking behavior provides a direct experimental route to reconstruct the underlying metric of the droplet, both in continuum Landau levels~\cite{fletcher2021geometric,crepel2023geometric,mukherjee2022crystallization,schine2016synthetic,lunt2024realization} and in lattice realizations such as the Harper–Hofstadter model~\cite{aidelsburger2015measuring,leonard2023realization,impertro2024local,impertro2025strongly}, where additional corrections were shown to lead to a renormalized, emergent metric.

Beyond serving as a geometric diagnostic, the present framework naturally connects to dynamical probes of geometric response functions. In particular, the ability to engineer metric-sensitive drives and to isolate chiral responses forms the basis for the measurement protocol developed in our accompanying work~\cite{sisterViscositypaper}, where such dichroic schemes are extended to access Hall viscosity. In that context, time-dependent quadrupolar perturbations effectively implement controlled deformations of the underlying metric, allowing one to extract the Berry curvature associated with area-preserving deformations. From this perspective, perfect elliptic dichroism can be viewed as the minimal manifestation of metric selectivity, while more general circular metric modulations provide access to the corresponding geometric response coefficients.

Looking ahead, an important direction concerns interacting quantum Hall systems, in particular fractional quantum Hall states. While the perfect elliptic dichroism reported here relies on holomorphicity properties of ideal Chern bands or Laughlin-like wavefunctions, extending these ideas to more general correlated phases could provide a powerful tool to probe their internal geometry and intrinsic orbital spin~\cite{wen2012modular,read2009non,read:rezayi:2011,fremling2014hall,cho2014geometry,arouca2022quantum,cappelli2018bulk}. In this context, it is interesting to contrast our single-particle, one-body driving protocol with recent approaches targeting collective spin-2 “graviton” modes~\cite{xavier2025chiral,bacciconi2026chiralgravitonmodesnonabelian}, which rely on carefully engineered modulations of interaction terms. Establishing connections between these complementary strategies may offer a unified view of geometric responses in strongly correlated topological phases, including non-Abelian states~\cite{read:rezayi:2011,Zhu_interface} where the underlying geometric structure is expected to be particularly rich.

More broadly, our results highlight how the geometry of both the probe and the quantum state can be coherently engineered and exploited to reveal intrinsic properties of topological matter. This interplay between probe geometry and quantum geometry is not restricted to two-dimensional systems, and it would be natural to extend these ideas to higher-dimensional settings~\cite{Price_4D,Karabali_4D,Blagoje_4D,Karabali_2023,FQH_higher_dimension}. Exploring such extensions may open new avenues to probe geometric response functions in synthetic quantum systems and engineered materials, further bridging topology, geometry, and dynamical spectroscopy.

\section*{Acknowledgments}
We acknowledge discussions with Leonardo Mazza, Blagoje Oblak, Martin Weitz, Arif Warsi Laskar, Franz Huybrechts, Jie Wang, Carolina Paiva and Ashvin Vishwanath. This research was financially supported by the ERC Grant LATIS, the FRS-FNRS (Belgium), the EOS project CHEQS, the Fondation ULB and the ANR PEPR
Grant QUTISYM ANR-23-PETQ-0002. B.~M. acknowledges support from the Security and Quantum Information Group (SQIG) in Instituto de Telecomunica\c{c}\~{o}es, Lisbon. This work is funded by FCT/MECI through national funds and when applicable co-funded EU funds under UID/50008: Instituto de Telecomunicações (IT). B.~M. further acknowledges the Scientific Employment Stimulus --- Individual Call (CEEC Individual) --- 2022.05522.CEECIND/CP1716/CT0001, with DOI: \href{https://doi.org/10.54499/2022.05522.CEECIND/CP1716/CT0001}{10.54499/2022.05522.CEECIND/CP1716/CT0001}. T.~O. acknowledges support from JSPS KAKENHI Grant Number JP24K00548, JST PRESTO Grant No. JPMJPR2353, and JST PRESTO Convergence Research Grant No. JPMJCR26XA.

\bibliography{bib.bib}

\begin{widetext}
\appendix
\section{$\Gamma^{\mathrm{int}}_{\pm}$ for arbitrary $\tau'$}
\label{sec: Gammapm for arbitrary tauprime}
In this appendix we compute $\Gamma^{\mathrm{int}}_{\pm}$ for arbitrary $\tau'$. Similar to what we did in Eq.~\eqref{eq: atau in terms of ai}, we can write
\begin{align}
\label{eq: atauprime in terms of atau}
\begin{bmatrix}
a_{\tau'} & a_{\tau'}^{\dagger}  
\end{bmatrix} &=\frac{1}{\sqrt{2\tau'_2 \hbar eB}}\begin{bmatrix}
\pi_1 & \pi_2    
\end{bmatrix}\begin{bmatrix}
1 & 1\\
\tau' & \tau'^{*}
\end{bmatrix} =   \frac{\sqrt{\tau_2}}{\sqrt{\tau'_2}}\begin{bmatrix}
a_\tau & a_\tau^\dagger    
\end{bmatrix}\begin{bmatrix}
1 & 1\\
\tau & \tau^{*}
\end{bmatrix}^{-1}\begin{bmatrix}
1 & 1\\
\tau' & \tau'^*
\end{bmatrix} \nonumber \\
&=\frac{1}{\sqrt{\tau_2\tau'_2}}\begin{bmatrix}
a_\tau & a_\tau^\dagger    
\end{bmatrix}\begin{bmatrix}
i(\tau^*-\tau') & i(\tau^*-\tau'^{*})\\
-i(\tau-\tau') & -i(\tau-\tau'^*)
\end{bmatrix},
\end{align}
and the same $SU(1,1)$ matrix relates $\begin{bmatrix}b_{\tau'}\ b^\dagger_{\tau'}   
\end{bmatrix}$ and $\begin{bmatrix}b_{\tau}\ b^\dagger_{\tau}   
\end{bmatrix}$. Using this we can show that
\begin{align}
\label{eq: ztauprime in terms of btau,btaudagger,atau,ataudagger}
z_{\tau'} &=\frac{\ell_{B}}{\sqrt{2\tau_2}}\left(i(\tau^*-\tau')b_{\tau}-i(\tau-\tau'^*)b_{\tau}^\dagger +i(\tau^* -\tau')a_{\tau} -i(\tau-\tau')a_{\tau}^\dagger\right) \nonumber, \\
z_{\tau'}^* &=\frac{\ell_{B}}{\sqrt{2\tau_2}}\left(-i(\tau-\tau'^*)b^\dagger_{\tau}+i(\tau^*-\tau')b_{\tau} -i(\tau -\tau'^*)a^\dagger_{\tau} +i(\tau^*-\tau'^*)a_{\tau}\right).
\end{align}
This allows us to compute the matrix elements explicitly,
\begin{align}
&\bra{\Psi_{n_2,l_2}(\tau)} z_{\tau'}\ket{\Psi_{n_1=0,l_1}(\tau)}=-i\frac{\ell_{B}}{\sqrt{2\tau_2}}(\tau-\tau')\delta_{n_2,1}\delta_{l_1,l_2} \text{ and }\nonumber \\
&\bra{\Psi_{n_2,l_1}(\tau)} z^*_{\tau}\ket{\Psi_{n_1=0,l_1}(\tau)}=-i\frac{\ell_{B}}{\sqrt{2\tau_2}}(\tau-\tau'^{*})\delta_{n_2,1}\delta_{l_1,l_2}, \text{ with } n_2\neq 0,   
\end{align}
and hence, using Eqs.~\eqref{eq: dichroic rates}, we find
\begin{align}
\Gamma^{\mathrm{int}}_{+}=\frac{2\pi E^2}{\hbar^2} l_{\mathrm{bulk}}\ell^2_{B} \frac{|\tau-\tau'|^2}{2\tau_2\tau_2'}=A\frac{E^2}{\hbar^2}\frac{|\tau-\tau'|^2}{2\tau_2\tau_2'} \quad \text{and} \quad \Gamma^{\mathrm{int}}_{-}=\frac{2\pi E^2}{\hbar^2} l_{\mathrm{bulk}}\ell^2_{B} \frac{|\tau-\tau'^{*}|^2}{2\tau_2\tau_2'}=A\frac{E^2}{\hbar^2}\frac{|\tau-\tau'^{*}|^2}{2\tau_2\tau_2'},
\end{align}
so that
\begin{align}
\frac{\Gamma^{\mathrm{int}}_{+}-\Gamma^{\mathrm{int}}_{-}}{2A}=\frac{E^2}{\hbar^2} \mathcal{C}.    
\end{align}
As announced, the difference is independent of $\tau'$, but only in the case $\tau'=\tau$, when the probe is adjusted to the shape dictated by $g^{ab}$, $\Gamma^{\mathrm{int}}_{+}=0$ since the LLL is an ideal Chern band.

\section{The Harper-Hofstadter model on a square lattice and its continuum limit}
\label{eq:continuum limit}
We first introduce magnetic translation operators; we use them to easily take the continuum limit of the Harper-Hofstadter Hamiltonian Eq.~\eqref{eq:hofst_latt}.

\subsection{Magnetic translations}
Using the kinetic momenta $\pi_a=p_a-eA_a$, which obey $[\pi_a,\pi_b]=i \frac{\hbar^2}{\ell_B^2} \epsilon_{ab}$,
we can introduce magnetic translation operators 
\begin{equation}
\label{eq:magnetic_translation}
T_{\bf{d}}=\exp\left(i \frac{d^a\pi_a}{\hbar}\right);
\end{equation}
As a consequence of the Baker-Campbell-Hausdorff relation it follows that
\begin{equation}
	\label{eq:magnetic_translation_commutator}
	T_{\mathbf d_1}T_{\mathbf{d}_2} =  T_{\mathbf d_1+\mathbf d_2} \exp\left(-i\,\frac{d_1^ad_2^b\epsilon_{ab}}{2\ell_B^2}\right).
\end{equation}

\subsection{The continuum limit of the Harper-Hofstadter model}
Consider the eigenstates $\ket{\psi_i}$ of the tight-binding Harper-Hofstadter Hamiltonian $H_{\rm HH}$ on a square lattice, with lattice spacing $a_0$; we decompose them over the basis given by the lattice sites ${\bf R}=(ma_0,na_0)$ as
$\ket{\psi_i}=\sum_{\bf R}C_{{\bf R},i}\ket{{\bf R}}$
\begin{equation}
    \begin{split}
    E_i \ket{\psi_i}&= H_{\rm HH} \ket{\psi_i} \\&=
    \sum_{\bf R} C_{{\bf R},i} \Biggl(-J_x \Bigl(\ket{{\bf R}+a_0\hat x}+\ket{{\bf R}-a_0\hat x}\Bigr)-J_y\Bigl(e^{i2\pi\alpha m}\ket{{\bf R}+a_0 \hat y}+e^{-i2\pi\alpha m}\ket{{\bf R}-a_0\hat y}\Bigr)\Biggr),
    \end{split}
\end{equation}
where $2\pi\alpha=a_0^2/\ell_B^2$ is the magnetic flux per plaquette.

The previous eigenvalue equation can be equivalently written as
\begin{equation}
    \begin{split}
    E_i C_{{\bf R},i} =& -J_x \Bigl(C_{{\bf R}-a_0\hat x,i}+C_{{\bf R}+a_0\hat x,i}\Bigr)-J_y\Bigl(e^{i2\pi\alpha m}C_{{\bf R}-a_0 \hat y,i}+e^{-i2\pi\alpha m}C_{{\bf R}+a_0 \hat y,i}\Bigr).
    \end{split}
\end{equation}
If we promote the (discrete) probability amplitudes $C_{{\bf R},i}$ to a family of continuous functions $C_{i}({\bf R})$, we can rewrite the right-hand side of the previous equation using the magnetic-translation operators
\begin{subequations}
\begin{equation}
	\label{eq:hh_magnetic_translations}
	H=-J_x\left(T_{(a_0,0)}^{}+T_{(a_0,0)}^\dagger\right)-J_y\left(T_{(0,a_0)}^{}+T_{(0,a_0)}^\dagger\right);
\end{equation}   
Schr\"odinger's equation then reads
\begin{equation}
    E_i C_{i}({\bf R})=H C_i({\bf R}).
\end{equation}
\end{subequations}

Notice how, if we define a square lattice through
\begin{equation}
    \label{eq:hh_gauge_choice}
	\ket{{\bf R}}=\ket{m,n}=T_{(a_0,0)}^m T_{(0,a_0)}^n\ket{0,0}
\end{equation} 
where $x= a m$, $y=a n$ are the lattice points, we recover the Harper-Hofstadter Hamiltonian when projecting Eq.~\eqref{eq:hh_magnetic_translations} onto these lattice sites.
\begin{subequations}
Indeed, acting with $T_{(a_0,0)}$ onto $\ket{m,n}$ gives
\begin{equation}
    \begin{split}
    	T_{(a_0,0)}\ket{m,n}=&\ket{m+1,n}    
    \end{split}
\end{equation}
while, using Eq.~\eqref{eq:magnetic_translation_commutator}, the action of $T_{(0,a_0)}$ onto $\ket{m,n}$ yields
\begin{equation}
    \begin{split}
        T_{(0,a_0)}\ket{m,n}&=e^{i2\pi\alpha m}\ket{m,n+1}.
    \end{split}
\end{equation}	
\end{subequations}
These identities directly lead to
\begin{equation}
	\label{eq:hh_hamiltonian_sm}
    \begin{split}
H_{\rm HH}=&\sum_{\mathbf{R},\mathbf{R}'}\bra{\mathbf{R}'}H\ket{\mathbf{R}}  c_{\mathbf{R}'}^\dagger c_{\mathbf{R}}^{}=\\=&-J_x \sum_{m,n} c_{m+1,n}^\dagger c_{m,n} -J_y \sum_{m,n} e^{i2\pi\alpha m}c_{m,n+1}^\dagger c_{m,n}+{\rm h.c.}
\end{split}
\end{equation}	
Notice how Eq.~\eqref{eq:hh_gauge_choice} is a gauge choice: it reflects into the Landau gauge of Eq.~\eqref{eq:hh_hamiltonian_sm}.

It is now easy to analyze the continuum limit of the Harper-Hofstadter model; 
indeed, when the lattice spacing is much smaller than the magnetic length $a_0\ll \ell_B$ one expects, at low-energies, the amplitudes $C_{{\bf R},i}$ to vary smoothly over the lattice spacing $a$; 
as a consequence, it makes sense to perform an ordered expansion of the magnetic-translation operators appearing in Eq.~\eqref{eq:hh_magnetic_translations},
retaining only the lowest order terms.
When $a_0/\ell_B\ll 1$ we get
\begin{equation}
    H \simeq -2(J_x+J_y) + \frac{1}{2}\Bigl(2J_x a_0^2\pi_x^2 + 2J_y a_0^2 \pi_y^2\Bigr)
\end{equation}
up to $\mathcal O(a_0^4/\ell_B^4)$ corrections. 
This is exactly (up to the constant energy shift $-2(J_x+J_y)$) the result Eq.~\eqref{eq:ll_limit} quoted in the main text.
It is not hard to recover the correction terms $\delta H$ of Eq.~\eqref{eq:perturbation lowest_order} by expanding the exponentials in Eq.~\eqref{eq:hh_magnetic_translations} to higher order in $a_0/\ell_B$.

\end{widetext}
\end{document}